# Comparison of Lift and Drag Modulation Control for Ice Giant Aerocapture Missions

Athul Pradeepkumar Girija [1],[**],[**]

[1]School of Aeronautics and Astronautics, Purdue University, West Lafayette, IN 47907, USA

**ABSTRACT**

Aerocapture is an orbit insertion technique which uses atmospheric drag from a single pass to decelerate a spacecraft. Compared to conventional propulsive insertion, aerocapture can impart large velocity changes to the spacecraft with almost no propellant. At the far reaches of the outer Solar System, the ice giants remain the last class of planets to be explored using orbiters. Their enormous heliocentric distance presents significant mission design challenges, particularly the large ΔV required for orbit insertion. This makes aerocapture an attractive method of orbit insertion, but also challenging due to the comparatively large navigation and atmospheric uncertainties. The present study performs a comparison of the lift and drag modulation control and their implications for future missions. Lift modulation provides nearly twice the entry corridor width as drag modulation, and can thus accommodate larger uncertainties. Lift modulation offers continuous control throughout the flight enabling it to adjust the trajectory in response to the actual density profile encountered. Drag modulation offers much more benign aero-thermal conditions compared to lift modulation. With drag modulation, there is no control authority after the drag skirt jettison making the vehicle more susceptible to exit state errors from density variations encountered after the jettison event.

---

[****] To whom correspondence should be addressed, E-mail: athulpg007@gmail.com

## I. INTRODUCTION

Aerocapture is an orbit insertion technique which uses atmospheric drag from a single pass to decelerate a spacecraft [1, 2]. Compared to conventional propulsive insertion, aerocapture can impart large velocity changes to the spacecraft with almost no propellant [3]. At the far reaches of the Solar System, the ice giants remain the last class of planets to be explored using orbiter spacecraft [4, 5, 6]. Their enormous heliocentric distance presents significant mission design challenges, particularly the large $\Delta V$ required for orbit insertion [7]. This makes aerocapture an attractive method of orbit insertion at the ice giants, Uranus and Neptune [8, 9]. Figure 1 shows an illustration of the aerocapture maneuver with the vehicle entering the atmosphere, reducing its energy, and then exiting the atmosphere. To accommodate the uncertainties in the navigated delivery state, atmospheric, and vehicle aerodynamics and exit the atmosphere with the desired exit state, it is necessary for the vehicle to have aerodynamic control authority during its flight [10]. If the vehicle enters steep and penetrates too deep into the atmosphere, it will bleed too much energy and may not exit. If the vehicle enters shallow, it may not bleed enough energy and may exit without getting captured. Aerodynamic control allows the vehicle to autonomously control the trajectory within the corridor and hence the energy depletion. A recent NASA study has highlighted the need for a comparison of lift and drag modulation control at Uranus and Neptune [11]. The present study uses the Aerocapture Mission Analysis Tool (AMAT) to perform a comparison of the lift and drag modulation control and their implications for ice giant aerocapture missions [12].

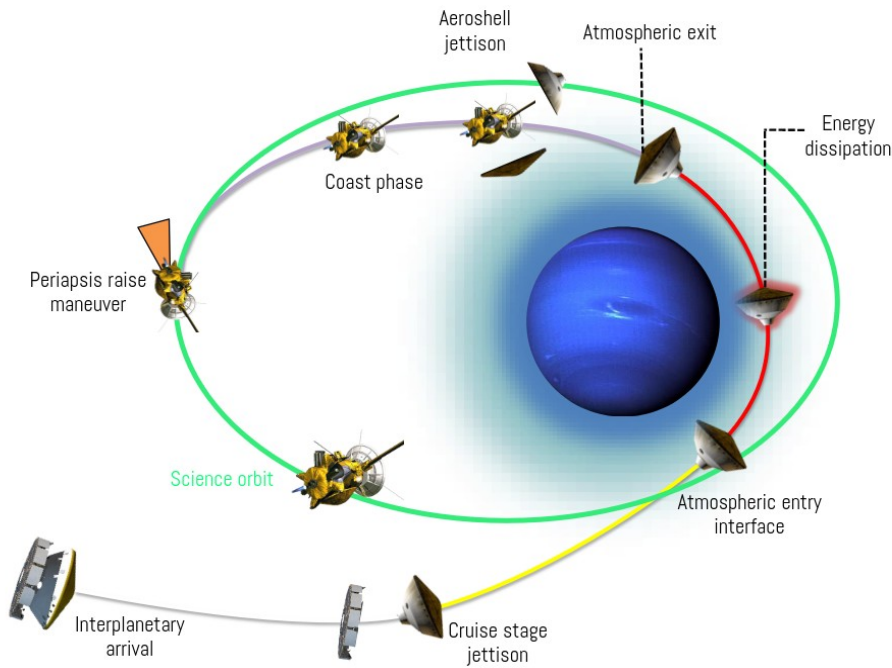

Figure 1. Schematic illustration of the aerocapture maneuver.

## II. LIFT MODULATION

Bank angle modulation (a subset of lift modulation) has been successfully used on the Apollo and MSL missions and is a proven and well understood technique. The only control variable is the bank angle, and by pointing the lift vector up or down, the vehicle can control its descent rate and energy depletion. Early studies of ice giant aerocapture at Neptune in the 2000s had used a mid-L/D (L/D=0.8) aeroshell to accommodate the large uncertainties [13]. However, recent studies have shown that by using high arrival v_inf trajectories, it becomes possible to use low-L/D aeroshells such as MSL (L/D = 0.24) while also enabling shorter flight times [14, 15]. Figure 2 shows the aerocapture trajectories for an MSL-like vehicle entering Uranus at 29 km/s. The target apoapsis is 500,000 km. The aerocapture corridor is [-12, -11] deg, with a width of 1.0 deg. The peak deceleration is in the range of 4–10g, and the peak heat rate is in the range of 1400–1800 W/cm$^2$. The peak heat rate for aerocapture is considerably less than that for entry probes which enter steeper, and is well within the tested limits of the HEEET thermal protection system [16]. The heat load is in the range of 200–300 kJ/cm$^2$, which is substantial but also expected to be within the capability of HEEET. Based on empirical relations, the TPS mass fraction is expected to be about 25%, and the structural mass fraction is also expected to be at 25%, leaving about 50% of the arrival mass to be inserted into orbit after aerocapture [17, 18].

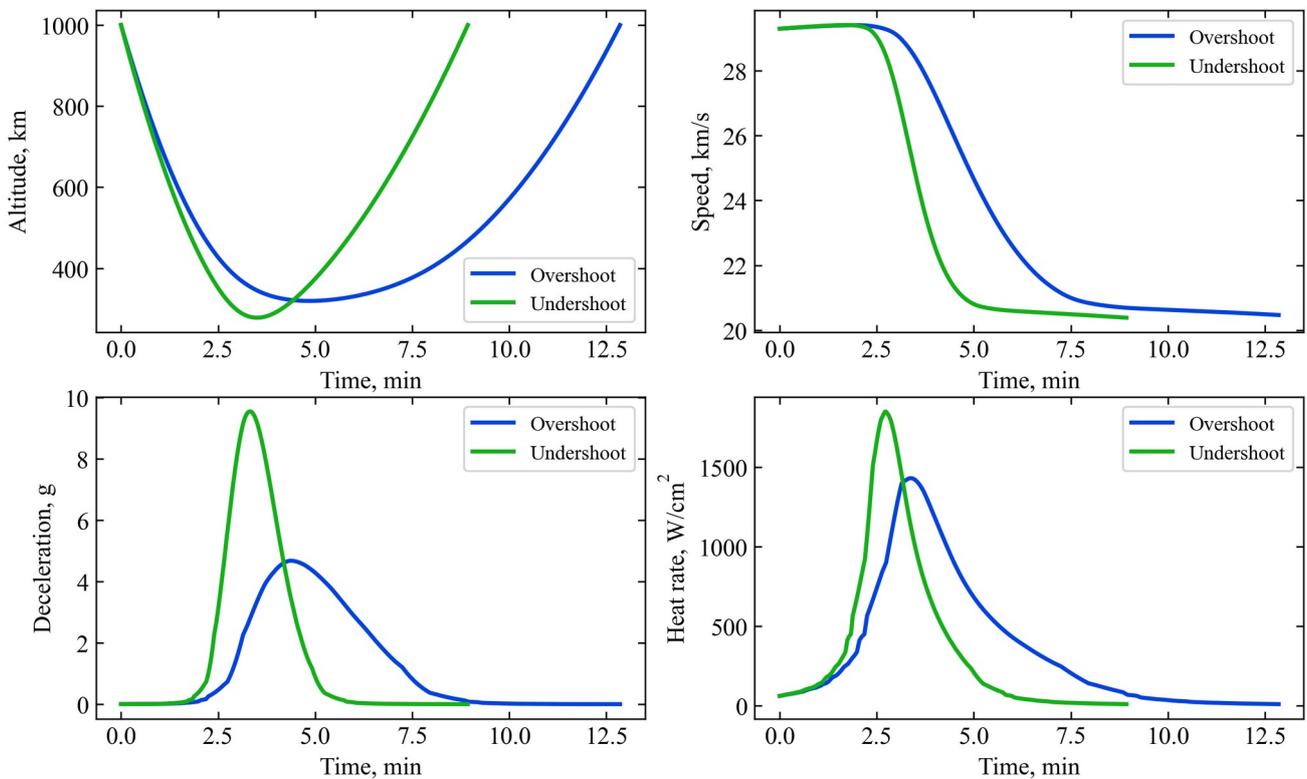

Figure 2. Lift modulation aerocapture trajectory at Uranus with an MSL-like aeroshell (L/D=0.24).

## III. DRAG MODULATION

Drag modulation is a simpler control technique that avoids the need for a propellant-fed reaction control thrusters required for bank angle modulation [19]. In its simplest variant, the single-event jettison, the only control variable is the time at which the drag skirt is jettisoned. By adjusting the jettison time, the energy depletion can be controlled. Unlike lift modulation which offers continuous control throughout the atmospheric flight, drag modulation provides no control authority after drag skirt jettison. Drag modulation uses a low ballistic coefficient entry system which enables much lower heating rates compared to lift modulation which uses a high ballistic coefficient rigid aeroshell. The low ballistic coefficient system decelerates much higher up in the atmosphere, keeping the heating rates low. However, the flexible TPS (such as carbon cloth used in ADEPT) can only accommodate smaller heat rates (200–300 $W/cm^2$), and thus cannot use high speed arrival trajectories [20, 21]. Figure 3 shows a nominal aerocapture trajectory for a 12-m ADEPT drag modulation vehicle (beta = 30 $kg/m^2$, BC ratio = 4.14) entering Uranus at 26 km/s [22]. The target apoapsis is 500,000 km. The aerocapture corridor is [-10.71, -10.25] deg, with a width of 0.46 deg. The peak deceleration is 5g, and the peak heat rate is in about 300 $W/cm^2$. The total heat load is about 77 $kJ/cm^2$. The estimated fraction of the arrival mass delivered to orbit with is 50%, which is the same as with lift modulation aerocapture [23].

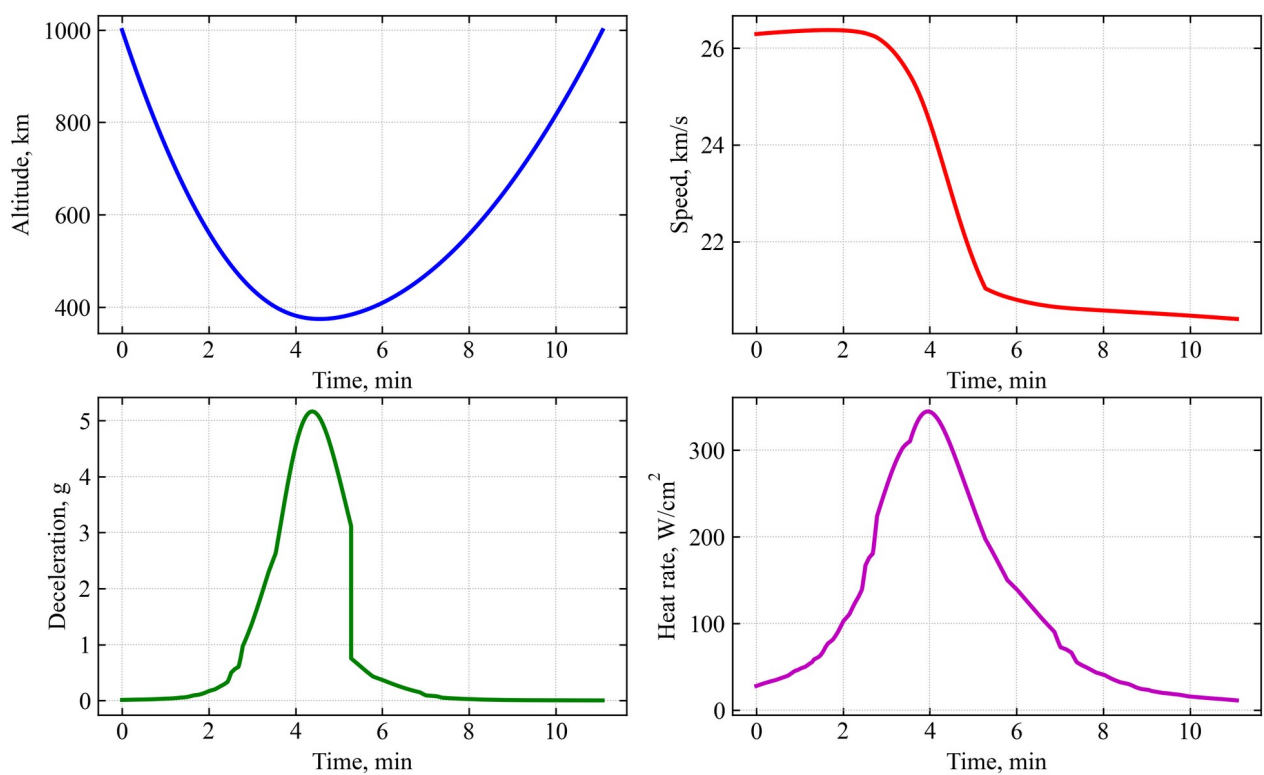

Figure 3. Drag modulation aerocapture trajectory at Uranus with a 12-m diameter ADEPT.

## IV. COMPARISON

Table 1 compares lift and drag modulation results at Uranus. The first observation is that lift modulation with the high entry speed provides corridor width that is nearly twice that of drag modulation. This implies lift modulation can accommodate higher navigation and delivery atmospheric uncertainties compared to drag modulation. In addition, lift modulation offers continuous control throughout the flight enabling it to adjust the trajectory in response to the actual density profile encountered. With drag modulation, there is no control authority after the drag skirt jettison making the vehicle susceptible to unexpected density pockets and other variations which may be present in the atmosphere [24].

The second difference is the peak heat rate which is in the range of 1400 – 1800 $W/cm^2$ for lift modulation, compared to 200–300 $W/cm^2$ for drag modulation aerocapture. The resulting total heat load is in the range of 200–300 $kJ/cm^2$ for lift modulation and 40–75 $kJ/cm^2$ for drag modulation. Hence the low ballistic coefficient system used in drag modulation offers a much more benign aero-thermal environment compared to lift modulation [25, 26].

The third difference is that for lift modulation, even with a high arrival speed, the peak heat rate is well within the tested limits for HEEET. For drag modulation, the peak heat rate is near the upper limit of the carbon cloth TPS which is tested to around 250 $W/cm^2$. Hence lift modulation architectures can accommodate high arrival speeds which can occur with high energy, short flight time trajectories, while drag modulation architectures tend to be more limited in terms of the maximum arrival speed due to the constraints on the peak heat rate of the carbon cloth TPS [27].

Table 1. Comparison of lift and drag modulation aerocapture at Uranus.

| Control Method | Corridor width, deg | TPS material | Peak heat rate, $W/cm^2$ | Total heat load, $kJ/cm^2$ | Delivered mass fraction, % |
|---|---|---|---|---|---|
| Lift Modulation | 1.00 | HEEET | 1400–1800 | 200–300 | 50 |
| Drag Modulation | 0.46 | Carbon cloth | 200–300 | 40–75 | 50 |

## V. CONCLUSIONS

The study compared lift and drag modulation control techniques and explored their implications for ice giant aerocapture. Lift modulation provides nearly twice the corridor width as drag modulation, and can thus accommodate larger delivery and atmospheric uncertainties. Lift modulation offers continuous control throughout the flight enabling it to adjust the trajectory in response to the actual density profile encountered. Drag modulation offers much more benign aero-thermal conditions for aerocapture compared to lift modulation. With drag modulation, there is no control authority after the drag skirt jettison making the vehicle more susceptible to off-nominal density variations.

## DATA AVAILABILITY

The results presented in the paper can be reproduced using the open-source Aerocapture Mission Analysis Tool (AMAT) v2.2.22. The data and code used to make the study results will be made available by the author upon request.